\documentstyle[prb,preprint,aps]{revtex}
\begin{document}
\draft

\title{Spin dynamics of SrCu$_2$O$_3$ and the Heisenberg ladder}

\author{Anders W. Sandvik,$^1$ Elbio Dagotto,$^1$ and
Douglas J. Scalapino$^2$}
\address{$^1$National High Magnetic Field Laboratory and Physics Department,
Florida State University, 1800 E. Paul Dirac Dr., Tallahassee, Florida 32306}
\address{$^2$Department of Physics, University of California, Santa Barbara,
California 93106}

\date{\today}

\maketitle

\begin{abstract}
The $S=1/2$ Heisenberg antiferromagnet in the ladder geometry is studied as
a model for the spin degrees of freedom of SrCu$_2$O$_3$.
The susceptibility and the spin echo decay rate are calculated
using a quantum Monte Carlo technique, and the spin-lattice relaxation rate
is obtained by maximum entropy analytic continuation of imaginary time
correlation functions. All calculated quantities are in reasonable agreement
with experimental results for SrCu$_2$O$_3$ if the exchange coupling
$J \approx 850$K, i.e. significantly smaller than in high-T$_c$ cuprates.
\end{abstract}

\vfill\eject

The Cu-O layers of SrCu$_2$O$_3$ have an internal structure of parallel double
chains (ladders).\cite{struct,azuma} Cu spins within a ladder are exchange
coupled with a strength expected to be comparable to that of high-T$_c$
cuprates, whereas the inter-ladder coupling is weak, arising from 90$^o$
Cu-O-Cu bonds. The spin degrees of freedom should therefore be well
described by the Heisenberg model on a single ladder,\cite{gopalan} defined
by the hamiltonian
\begin{equation}
\hat H = J_1\sum\limits_{i}\sum\limits_{a=1,2}
\vec S_{a,i} \cdot \vec S_{a,i+1} +
J_2 \sum\limits_{i} \vec S_{1,i} \cdot \vec S_{2,i},
\label{heisenberg}
\end{equation}
where $\vec S_{a,i}$ is a spin-1/2 operator at site $i$ of chain $a$.
It is now well established that this system has a gap between
the ground state and the lowest excitation for any ratio
$J_2/J_1\not=0$. For $J_1=J_2=J$, the gap is
$\Delta = 0.504J$.\cite{dagotto}

Recent experiments on SrCu$_2$O$_3$ have been carried
out by Azuma {\it et al.} \cite{azuma} and Ishida {\it et al.}.\cite{ishida}
Their results for the spin susceptibility $\chi$ and the $^{63}$Cu NMR
spin-lattice relaxation rate $1/T_1$ show clear evidence of a gap.
Accordingly, the spin-echo decay $1/T_{2G}$ rate saturates at low
temperatures, indicating a finite correlation length in the ground
state. \cite{ishida} However, comparing the data for $\chi$ and $1/T_1$
with theoretical low-temperature results for the Heisenberg ladder obtained by
Troyer {\it et al.},\cite{troyer} there is a significant discrepancy; $\chi$
indicates a gap  $\Delta \approx 420K$, whereas the behavior of $1/T_1$
suggests a gap close to $700$K.\cite{azuma,ishida,troyer} At first
sight, one would tend to believe that the gap extracted from $1/T_1$ is
the correct one, since the corresponding value of $J\approx 2\Delta$ is
then close to the exchange constants typically found in planar cuprates.

We have carried out quantum Monte Carlo (QMC) simulations of the
Heisenberg ladder, and obtained results for the quantities discussed
above. Here we present comparisons with the  experimental results, and
discuss a possible reason for the gap-size discrepancy found in earlier
work. We argue that the formula used to extract the gap from $1/T_1$ is
not applicable in the temperature regime where it was used,
and that the gap obtained from $\chi$ is more accurate. The
calculated $1/T_{2G}$ is also in close agreement with the experimental
result for $J \approx 850$K, corresponding to the smaller gap.

Troyer {\it et al.} calculated $\chi$ and $1/T_1$ for the ladder
by considering the magnon dispersions obtained in the limit
$J_2 \gg J_1$.\cite{troyer} The lowest branch is a single-magnon
state which is odd with respect to interchange of the two
chains ($k_y=\pi$). This remains the lowest excitation also
when $J_2 = J_1$. The smallest gap ($\Delta$) is at momentum $k_x=\pi$
along the chains. As $k_x\to 0$, the one-magnon branch crosses into
a multi-magnon continuum. At $k_x = 0$ the gap is $\approx 2\Delta$,
corresponding to a two-magnon excitation. At low temperatures the
thermodynamics of the ladder is thus obtained by populating the modes
with $k_x \approx \pi$, $k_y=\pi$. The susceptibility then has
the form\cite{troyer}
\begin{equation}
\chi \sim T^{-1/2} \hbox{e}^{-\Delta /T} .
\label{susc}
\end{equation}
For $T$ up to $\approx \Delta$ this form is in good agreement with results
from exact diagonalizations of small systems,\cite{barnes} as well as
quantum transfer matrix results.\cite{troyer} As mentioned above, the
agreement with experimental results for SrCu$_2$O$_3$ is also good,
with a $\Delta \approx 420$K.\cite{azuma}

The NMR spin-lattice relaxation rate is related to the dynamic
structure factor $S(\vec q,\omega)$ according to \cite{moriya}
\begin{equation}
1/ T_1 =
\hbox{$2 \over \hbar $} \sum\limits_{\vec q} |A_{\vec q}|^2
S(\vec q, \omega \to 0),
\label{t1moriya}
\end{equation}
where $A_{\vec q}$ is the nuclear hyperfine form factor.
At very low temperatures, the main contributions to $1/T_1$ come
from momentum transfers $q_x\approx 0, q_y=0$, i.e. both the initial
and final states are on the one-magnon branch at $k_x \approx \pi$.
Taking into account only these processes, Troyer {\it et al.}
obtained the leading low-temperature form \cite{troyer}
\begin{equation}
1/ T_1 \sim |A_{q=0}|^2 \hbox{e}^{-\Delta/T} .
\label{t1}
\end{equation}
A behavior close to exponential is seen for SrCu$_2$O$_3$ in the temperature
regime $100K \alt T \alt 300K$.\cite{azuma,ishida} At lower temperatures
$1/T_1$ is dominated by impurity effects. The $J\approx 1300$K extracted
from fits of (\ref{t1}) to experimental data is markedly different
form the $J \approx 850$K obtained  from the susceptibility.

One could certainly argue that SrCu$_2$O$_3$ is not a perfect ladder
system. Most likely, $J_1$ is not exactly equal to $J_2$. However, the
above theoretical forms only depend on the gap, and the disagreement
between the gaps from $\chi$ and $1/T_1$ cannot be explained by
$J_1\not= J_2$ alone. Furthermore, the coupling between the ladders
is expected to be weak. \cite{gopalan} Thus, before
discarding the single ladder as a good approximation
of the system, it is important to investigate its behavior in more
detail. Whereas the low-temperature form (\ref{susc}) for the
susceptibility has been verified to be accurate by comparisons with
numerical results, \cite{troyer,barnes} the form (\ref{t1}) for the
spin-lattice relaxation rate has not been tested numerically. At very
low $T$ it is hard to see why (\ref{t1}) should not apply. However,
the temperatures for which the fit to the experimental results were made
are not very low on the scale set by the gap. It is clear that there
will be large contributions to $1/T_1$
from processes with $q_x \approx \pi , q_y=\pi$ between the one-magnon
branch and the continuum at $k_x \approx 0$ if the temperature is
high enough for states at energies $\agt 2\Delta$ to be populated.
These processes are particularly important because the ladder has strong
short-range antiferromagnetic correlations. The matrix elements
entering the $q_x\approx \pi,q_y=\pi$ processes are therefore
much larger than those for $q_x\approx 0,q_y=0$. Hence, although the
$q_x\approx 0,q_y=0$ contributions are the only ones surviving
in the $T \to 0$ limit, it is quite likely that $1/T_1$ is actually
dominated by other processes at the upper range of temperatures
considered in the experiments.

We have calculated $1/T_1$ using the maximum entropy (ME) method \cite{maxent}
to analytically continue imaginary time correlation functions obtained
by a QMC technique. The spin-echo decay rate $1/T_{2G}$ is
related to the static susceptibility, \cite{pennington} which can be
calculated directly. We have used a recently developed QMC method based on
stochastic series expansion, \cite{qmc} which produces results free from
systematical errors associated with Trotter based methods.

The calculations of $1/T_1$ and $1/T_{2G}$ require knowledge of the
Cu nuclear hyperfine interactions. For the high-Tc cuprates
La$_{2-x}$Sr$_x$CuO$_4$ and  YBa$_2$Cu$_3$O$_{6+x}$ the hyperfine couplings
are well described by the Mila-Rice form, \cite{hyper} with an anisotropic
on-site coupling with components  $A_\perp$ and $A_\parallel$, and an
isotropic nearest-neighbor transferred coupling $B$. Typical values reported
are $B=$41kOe$/\mu_B$, $B/A_\perp =1.2$,
and $B/A_\parallel = -0.25$.\cite{hyper} Knight shift
measurements on SrCu$_2$O$_3$ by Ishida {\it et al.} \cite{ishida}
indicate that $B$ is much smaller in this compound. Assuming that a
single ladder picture is appropriate for the spin system as well as
for the hyperfine couplings, and that the $B$-couplings have equal
strengths along and across the chains, the Knight shift results\cite{ishida}
give the relations
\begin{mathletters}
\begin{eqnarray}
A_\perp + 3B_\perp & = & {\rm 48kOe/\mu_B}, \label{rela} \\
A_\parallel + 3B_\parallel & = & {\rm -120kOe/\mu_B}, \label{relb}
\end{eqnarray}
\label{relations}
\end{mathletters}
where we have not excluded an anisotropic $B$. Assuming that the
on-site couplings remain close to their standard Mila-Rice
values, the transferred couplings in SrCu$_2$O$_3$ are thus
$B_\perp \approx 4{\rm kOe/\mu_B}$, and $B_\parallel \approx
15{\rm kOe/\mu_B}$. Considering experimental uncertainties, these
estimates are probably consistent with $B_\perp = B_\parallel$.
In any case, the magnitude of $B$ is much smaller than the typical
two-dimensional cuprate value.

Below we present results for $\chi$, $1/T_1$ and $1/T_{2G}$. For the
NMR rates we use the relations (\ref{relations}) and several values of
the ratios $B_\perp /A_\perp$ and $B_\parallel /A_\parallel$.
We consider two choices for the exchange $J$, corresponding
to approximately the values obtained before from $\chi$ and $1/T_1$;
$J=850$K and $J=1200$K.

Fig.~\ref{figsus} shows our results for the spin susceptibility along with
the experimental results by Azuma {\it et al.}. \cite{azuma} We have
used lattices with up to $N=2\times 128$ spins, which for the temperatures
considered here is enough for finite-size effects to be negligible.
The agreement with earlier numerical results \cite{barnes,troyer} is
good. With a g-factor $g=2$ the QMC results for $J=850$K agree well with
the experimental data. For $J=1200$K the resulting $\chi$ fails to
reproduce the experimental results. Even if
$g$ is adjusted it is not possible to obtain a reasonable agreement for
any sizeable temperature regime. If indeed $J$ is $1200$K or larger,
this discrepancy is hard to explain.

For extracing $1/T_1$ we have calculated the $r$-space imaginary time
correlation functions corresponding to Eq.~(\ref{t1moriya}), and
continued these
numerically to real frequencies using the ME technique. \cite{maxent}
We have obtained the relevant correlation functions to within relative
statistical errors  of $10^{-4} - 10^{-3}$ for systems with up to
$2 \times 128$ spins. Even with this high accuracy the continued functions
have some uncertainties. At high temperatures the procedures can be tested
against exact diagonalization results, since the distribution of
$\delta$-functions that represent  the dynamic structure factor of a
small system then is dense enough that a small
broadening produces a smooth function, which can be compared with the
results obtained with the ME technique. As the temperature is lowered,
the number of $\delta$-functions with significant weight decreases
rapidly. For the largest systems that can be exactly diagonalized the
presence of many gaps then prohibit meaningful comparisons with ME
results, since this method cannot resolve structure on that scale.
At temperatures where comparisons are meaningful, the ME method produces
results in good agreement with exact results for a 16 site Heisenberg
chain.\cite{rates1d} Additional evidence that this is a reliable
method for obtaining $1/T_1$ stems from work on the two-dimensional
Heisenberg model,\cite{rates2d} where good agreement with experiments
on La$_2$CuO$_4$ was found, as well as results for the one-dimensional
Heisenberg model, \cite{rates1d} which exhibit the behavior
expected on theoretical grounds.

For the ladder, results obtained using the ME technique become
uncertain at temperatures where the gap opens up, and the weight
for $\omega \approx 0$ relative to the weight for $\omega > \Delta$
decreases rapidly. We believe that our results are accurate
for $T \agt \Delta/2$, and become increasingly inaccurate for lower
$T$. Here we present results for $T/J \ge 0.2$. The accuracy of the
results are probably not higher than tens of percent in the
worst cases. Nevertheless, they are useful for establishing the
general trends.

Fig.~\ref{figt1} shows results for several values of the ratio
$B_\perp /A_\perp$, with relation (\ref{rela}) satisfied. Interestingly,
for a strictly local interaction ($B_\perp /A_\perp=0$) and $J=1200$K
there is very good agreement with the experiment. However, in this
case (\ref{rela}) gives $A_\perp \approx {\rm 48kOe/\mu_B}$, which
is much higher than one would expect. It is believed that the on-site
couplings should be less sensitive to details of the
structure of a particular material than the transferred couplings,
and therefore one expects $A_\perp \approx {\rm 34kOe/\mu_B}$ as in
planar cuprates. \cite{hyper} For $J=850$K the best over-all agreement
is obtained with $B_\perp /A_\perp \approx 0.1$, which gives a
reasonable value for $A_\perp$ as well. However, the slope of the curve
is different from the experimental one. Nevertheless, it is interesting
to note that the magnitude of $1/T_1$ agrees with the experimental
curve to within a factor of 2 in the regime $150K \alt T \alt 300K$,
with a $J=850$K that accounts for the susceptibility as well.

A clear indication that $1/T_1$ in the regime considered here is
not dominated by $\vec q \approx 0$ processes is that there is a
significant decrease in $1/T_1$ with increasing $B_\perp /A_\perp$.
With (\ref{rela}) satisfied, the form factor $A_{q=0}$ remains
constant, and hence the low-T form (\ref{t1})
predicts a $1/T_1$ that does not change with $B_\perp /A_\perp$.
As we argued above, one can expect processes with
$\vec q \approx (\pi,\pi)$ to be important at these temperatures,
and the decrease in $1/T_1$ with increasing $B_\perp /A_\perp$
is then naturally explained by the decrease in the form factor
at $\vec q=(\pi,\pi)$.

Only rough estimates of the behavior of the spin-echo decay rate
of the Heisenberg ladder have been made.\cite{ishida}
It is dominated by the indirect
nuclear spin-spin interactions induced by the coupling to the electronic
spin system. Pennington and Slichter derived the form \cite{pennington}
\begin{equation}
{1\over T_{2G}} =
\Bigl [{0.69\over 2\hbar^2} \sum\limits_{x \not= 0}
J_z^2 (0,\vec x) \Bigr ] ^{1/2},
\end{equation}
where $J_z(\vec x_1,\vec x_2)$ is the $z$-component of the induced
interaction between nuclei at $\vec x_1$ and $\vec x_2$:
\begin{equation}
J_z(\vec x_1,\vec x_2) = -\hbox{$1\over 2$}
\sum\limits_{i,j} A(\vec x_1 - \vec r_i)A(\vec x_2 - \vec r_j)\chi (i-j),
\end{equation}
and $0.69$ is the natural abundance of $^{63}$Cu isotope.
The only non-zero hyperfine couplings are $A(0)=A_\parallel$
and $A(1)=B_\parallel$. For a system with a gap, the static susceptibility
$\chi (i-j) = \int_0^\beta d\tau \langle S^z_i (\tau ) S^z_j (0) \rangle$
decays exponentially with $|\vec r_i-\vec r_j|$ even at $T=0$.
$1/T_{2G}$ calculated for a ladder with $2 \times 128$ spins at
$T/J \ll \Delta$ is therefore a good approximation to the $T=0$ result of an
infinite system. For this quantity we can thus obtain ground
state as well as finite-$T$ results.

Fig.~\ref{figt2} shows results obtained using relation (\ref{relb})
and several ratios $B_\parallel /A_\parallel$. If $A_\parallel$ is to
remain close to its value in planar cuprates we need
$B_\parallel /A_\parallel \approx -0.1$, which with $J=850$K
indeed gives a quite good agreement with the experimental result.
An almost perfect agreement is obtained with $J=850$K and
$B_\parallel /A_\parallel \approx -0.12$. With $J=1200$K a slightly
larger $B_\parallel /A_\parallel$ is needed to produce an approximate
agreement with the experiment, but the slope of the numerical curve cannot
be reproduced as well as with $J=850$K. Note that for a strictly
local coupling ($B_\parallel=0$) and $J=1200$K, which
gave a good agreement for $1/T_1$ (Fig.~\ref{figt1}), $1/T_{2G}$ is
almost an order of magnitude too small.

All the above results were obtained with the assumption that the
chain coupling $J_1$ is equal to the rung coupling $J_2$. It is
important to consider also the more general case of non-equal couplings.
Allowing $J_2 \not = J_1$ we find that the best agreement with the
susceptibility is obtained with $J_2/J_1 \approx 0.8$,
and $J_1 \approx 1100$K (this requires a g-factor $g \approx 2.1$).
The results for $1/T_1$ and $1/T_{2G}$
calculated with these parameters are not in significantly better
agreement with the experiments than those shown in
Figs.~\ref{figt1} and \ref{figt2}, however.

We conclude that the experimentally measured $\chi$ and $1/T_{2G}$
for SrCu$_2$O$_3$ can be well accounted for by a Heisenberg ladder
with $J=850$K, and the experimentally determined hyperfine couplings.
The calculated $1/T_1$ agrees with the experiment to within a factor of
$2$. The reason for the discrepancies in this quantity
could be details of the hyperfine
couplings not taken into account here, such as possible
differences in the transferred couplings $B$ along a chain and on a rung.
$1/T_1$ is a direct measure of the low-frequency spin fluctuation
spectral weight, whereas $1/T_{2G}$ is given by a frequency integral.
It is therefore likely that $1/T_1$ is more sensitive than $1/T_{2G}$ to
slight deviations from the assumed hyperfine relations (\ref{relations})
in the regime where the low frequence spin fluctuation spectral
weight drops rapidly. Hence, we consider the agreement with the
experiment to within a factor 2 reasonable. We propose that the
reason for the discrepancies reported earlier\cite{azuma,troyer}
for the gaps extracted from $\chi$ and $1/T_1$ is that contributions
to $1/T_1$ arising from processes with momentum transfer
$q_x \approx \pi , q_y=\pi$ are important at high temperatures.
Since only a narrow range of relatively high temperatures is accessible
experimentally, a fit to the low-$T$ form (\ref{t1}) can give misleading
results for $\Delta$.

The value of $J$ hence appears to be smaller than the
typical values observed in high-T$_c$ cuprates. This is puzzling,
since the Cu-O bond structure of the SrCu$_2$O$_3$ ladders is the same
as that of the two-dimensional cuprates.\cite{struct,azuma}
One possible explanation
for the reduced value is that $J$ represents an effective coupling once
inter-ladder effects are taken into account. The weak frustrated
ferromagnetic coupling between ladders is expected to enhance the
gap,\cite{gopalan} and is therefore not a likely mechanism for reducing
the effective $J$. On the other hand, a $c$-axis coupling
reduces the gap, and may be important in SrCu$_2$O$_3$. However,
preliminary QMC results for the susceptibility of a
stack of weakly coupled ladders with $J > 1000$K do not compare as
favourably with the experiments as the single ladder result with
$J=850$K shown in Fig~\ref{figsus}. We would also like to point out the
possibility that the Madelung potentials of SrCu$_2$O$_3$ and the
two-dimensional cuprates might be different. This would alter the size of
the energy denominators associated with the superexchange interaction.

We would like to thank M. Takano, Y. Kitaoka, and  co-workers for
providing their experimental data. This work is supported by the
Office of Naval Research under Grant No. ONR N00014-93-0495 (A.W.S. and
E.D.) and the Department of Energy under Grant No. DE-FG03-85ER45197 (D.J.S.).

\begin{figure}
\caption{QMC results for the spin susceptibility of
the Heisenberg ladder with $J=850$K and $J=1200K$ compared with
the experimental results for SrCu$_2$O$_3$. A g-factor $g=2$ was
used for both sets of numerical results.}
\label{figsus}
\end{figure}

\begin{figure}
\caption{The spin-lattice relaxation rate calculated using QMC and
ME compared with the experimental results by Ishida
{\it et al.}\protect{\cite{ishida}} (thick solid curves).
The upper and lower panels show results for $J=850$K and $J=1200$K,
respectively.
The hyperfine couplings used satisfy (\protect{\ref{rela}}). The ratios
$B_\perp /A_\perp$ are $0$ (open circles) $0.05$ (solid circles),
$0.10$ (open squares), and $0.20$ (solid squares).}
\label{figt1}
\end{figure}

\begin{figure}
\caption{QMC results for the spin-echo decay rate compared with
the experimental results by Ishida {\it et al.} \protect{\cite{ishida}}
(solid curves).  The upper and lower panels show results for $J=850$K
and $J=1200$K, respectively. The hyperfine couplings used satisfy the relation
(\protect{\ref{relb}}). The ratios $B_\parallel /A_\parallel$
are $0$ (open circles) $-0.05$ (solid circles), $-0.10$ (open squares),
and $-0.15$ (solid squares). The dashed curve is the best fit for
$J=850$K, with $B_\parallel /A_\parallel=-0.12$.}
\label{figt2}
\end{figure}

\end{document}